\documentclass[referee]{aa}
\usepackage{graphicx}
\usepackage{txfonts}
\begin{document}


\title{Are collisions with neutral hydrogen important for modelling the Second Solar Spectrum of Ti {\sc i} and Ca {\sc ii} ?
}
\author{M. Derouich, J. Trujillo Bueno\thanks{Consejo Superior de Investigaciones Cient\'\i ficas (Spain)} and R. Manso Sainz}
\institute{Instituto de Astrof\'\i sica de Canarias, E-38205 La Laguna,
Tenerife, Spain
}

\titlerunning{Influence of isotropic collisions on the scattering line polarization}
\authorrunning{Derouich et al.}

\date{Received XXXX / XXXX}

\abstract
{The physical interpretation of scattering line polarization offers a novel diagnostic window for exploring the thermal and magnetic structure of the quiet regions of the solar atmosphere.
} 
{To evaluate the impact of isotropic collisions with neutral hydrogen atoms on the scattering polarization signals of the 13 lines of multiplet 42 of \ion{Ti}{i} and on those of the K line and of the IR triplet of \ion{Ca}{ii}, with emphasis on the collisional transfer rates between nearby $J$-levels.
}
{We calculate the linear polarization produced by scattering processes
in a plane-parallel layer illuminated by the radiation field from the underlying solar photosphere. We consider realistic multilevel models and solve the statistical equilibrium equations for the multipolar components of the atomic density matrix.
}
{We give suitable formulae for calculating the collisional rates as a function of temperature and hydrogen number density. We confirm that the lower levels of the 13 lines of multiplet 42 of \ion{Ti}{i} are completely depolarized by elastic collisions. Upper-level collisional depolarization turns out to have an unnoticeable impact on the emergent linear polarization amplitudes, except for the ${\lambda}4536$ line for which it is easier to notice a rather small depolarization caused by the collisional transfer rates. Concerning the \ion{Ca}{ii} lines, we show that the collisional rates play no role on the polarization of the upper level of the K line, while they have a rather small depolarizing effect on the atomic polarization of the metastable lower levels of the \ion{Ca}{ii} IR triplet.
}
{Although the collisional transfer rates between nearby $J$-levels seem to play a minor role for most of the lines we have considered in this paper, except for the magnetically insensitive ${\lambda}4536$ line of \ion{Ti}{i}, they might be important for other atomic or molecular systems with closer $J$-levels (e.g., hyperfined structured multiplets and/or molecules). Therefore, future research in this direction will be worthwhile.
}
 
\maketitle
\section{Introduction}

The physical interpretation of observations of the spectral line  polarization produced by scattering processes offers a unique opportunity to learn about the thermal and magnetic properties of the ``quiet" regions of the solar atmosphere (e.g., the recent reviews by Stenflo 2003 and Trujillo Bueno 2003). Of great scientific interest is the possibility of obtaining empirical information on hidden aspects of solar magnetism, such as tangled magnetic fields in the solar photosphere (e.g., Trujillo Bueno et al. 2004; Stenflo 2004; Manso Sainz et al. 2004), and/or very weak magnetic fields in the solar chromosphere (e.g., Manso Sainz \& Trujillo Bueno 2001, 2003, 2007; Trujillo Bueno \& Manso Sainz 2002; Holzreuter et al. 2006). In order to be able to achieve such goals in a reliable way it is crucial to carry out detailed radiative transfer calculations using sufficiently accurate values of the rates of isotropic collisions with neutral hydrogen atoms, such as those computed by Derouich et al. (2003; 2004; 2005$a,
 b$) within the framework of the semi-classical theory of collisions. The present paper addresses this issue by investigating their influence on the scattering polarization of some spectral lines of the linearly polarized solar limb spectrum, which are of great diagnostic value. 

Collisions with neutral hydrogen atoms in the solar atmosphere may modify the atomic level polarization that anisotropic radiation pumping processes induce in the atomic and molecular energy levels. We have elastic collisional rates, $D^{(K)}(J)$, which normally tend to destroy the population imbalances between the magnetic sublevels pertaining to each level of total angular momentum $J$. In addition, we may also have significant transfer rates between different $J$ levels
due to inelastic and super-elastic collisions, $C^{(K)}_I(J,J_l)$ and $C^{(K)}_S(J,J_u)$, respectively. These collisional transfer rates may modify the population and/or the atomic polarization of the level $J$ under consideration through collisional transfer from the lower ($J_l$) and upper ($J_u$) levels. Here we pay particular attention to both type of collisions with neutral hydrogen atoms, with emphasis on the (typically neglected) collisional transfer rates between the $J$ levels pertaining to each spectral term of the chosen multilevel model.

In this paper we focus our attention on the 13 lines of multiplet No. 42 of Ti {\sc i} and on the following lines of Ca {\sc ii}: the K line at 3934 \AA\ and the IR triplet at 8498 \AA, 8542 \AA\ and 8662 \AA. The reason for choosing multiplet 42 of Ti {\sc i}, which results from the transitions between the upper term $y^5F^0$ and the (metastable) lower term $a^5F$, is because one of the lines (i.e., ${\lambda}4536$) is insensitive to magnetic fields (the Land\'e factors of its lower and upper levels are zero), while the remaining 12 lines are sensitive to the Hanle effect. Therefore, ratios of the observed polarization amplitudes in suitably chosen line pairs can be used to infer the presence of tangled magnetic fields in the quiet solar atmosphere (Manso Sainz et al. 2004). The scientific motivation for choosing the above-mentioned lines of Ca {\sc ii} is because they are of great diagnostic interest for investigating the 
thermal and magnetic structure of the quiet solar 
chromosphere (Manso Sainz \& Trujillo Bueno 2001, 2003, 2007; Trujillo Bueno \& Manso Sainz 2002; Holzreuter et al. 2006). 

\section{Formulation of the problem}
 
We consider a plane-parallel layer, of Ti {\sc i} atoms or Ca {\sc ii} ions in the absence of magnetic fields, illuminated by the radiation field of the underlying solar photosphere. We assume that this anisotropic radiation field is axisymmetric with respect to the local vertical and unpolarized. Therefore, the following spherical tensors components suffice to describe the optical pumping properties of the incident radiation field (e.g., Landi Degl'Innocenti \& Landolfi 2004)

\begin{equation}
  J^0_0=\,\frac{1}{2}\int_{-1}^1I(\mu)  {\rm d}\mu ,     
\end{equation}
\begin{equation}
  J^2_0=\,\frac{1}{4\sqrt{2}}\int_{-1}^1 
	[(3\mu{}^2-1)I(\mu)] {\rm d}\mu ,       
\end{equation}
where $I(\mu)$ is the intensity and $\mu$ the cosine of the angle that
the radiation beam under consideration forms with the local vertical. In the reference system where the $z$-axis (i.e., the quantization axis of total angular momentum) is chosen along the symmetry axis of the incident radiation field (i.e., along the solar local vertical direction), the anisotropic illumination of the atoms produce population imbalances which we quantify by the following multipolar components of the atomic density matrix:

\begin{equation}
  \rho^K_0(J)\,=\,\sum_{M=-J}^{J}(-1)^{J-M}\sqrt{2K+1}\left(
	\begin{array}{ccc}
	J & J & K \\
	M & -M & 0 
	\end{array}\right)\,{\rho_J(M,M)},
\end{equation}
where ${\rho_J(M,M)}$ is the population of the substate with magnetic quantum number $M$, while 
$K=0, 2, ..., 2J$ for the Ti {\sc i} levels (Ti {\sc i} levels have integer total angular momentum values) and
$K=0, 2, ..., 2J-1$ for the Ca {\sc ii} levels (Ca {\sc ii} levels have half-integer total angular momentum values). Note that $\rho^0_0(J)$ 
is proportional to the total population of the $J$ level under consideration, while $\rho^2_0(J)$ is the so-called alignment of the level (which quantifies the degree of population imbalances among its magnetic sublevels). Note also that we are safely neglecting the hyperfine structure of the odd isotopes, given that the most abundant isotopes have zero nuclear spin ($87\%$ for titanium and practically $100\%$ for calcium).

Both, radiative and collisional transitions contribute to the $\rho^K_0(J)$ values corresponding to each $J$ level of the assumed atomic model. In the regime of the impact approximation the rate of change of $\rho^K_0(J)$ can be written as
\begin{equation}
{d\over{dt}}\rho^K_0(J)\,=\,[{d\over{dt}}\rho^K_0(J)]_{\rm rad}\,+\,[{d\over{dt}}\rho^K_0(J)]_{\rm coll}.
\end{equation}

The contribution of the radiative rates has been studied in detail elsewhere (e.g., Manso Sainz \& Landi Degl'Innocenti 2002; Manso Sainz et al. 2006) and its importance will not be emphasized further here. Adopting the same notation of the monograph by Landi Degl'Innocenti \& Landolfi (2004) we write the contribution of collisional transitions as follows:

\begin{eqnarray} 
\big[\frac{d}{dt}\; \rho_0^{K} (\alpha J)\big]_{\rm coll} & = &  
  - \big[ \sum_{{\alpha}_lJ_l}  C^{(0)}_S({\alpha}_l  J_l, \alpha J) \nonumber \\ && +\sum_{{\alpha}_u J_u} C^{(0)}_I({\alpha}_u  J_u, \alpha J) + D^{(K)}(\alpha J) \big] \; \rho_0^{K} (\alpha J)  \\
&& + \sum_{{\alpha}_l J_l}  \sqrt{ \frac{2J_l+1} {2J+1}}
C_I^{(K)}(\alpha J, {\alpha}_l J_l) \;  \rho_0^{K} ({\alpha}_l  J_l)   \nonumber \\ && + \sum_{{\alpha}_u J_u }  
\sqrt{ \frac{2J_u+1} {2J+1}}
C_S^{(K)}(\alpha J, {\alpha}_u J_u) \;  \rho_0^{K}({\alpha}_u  J_u), \nonumber 
\end{eqnarray}
where $D^{(K)}(\alpha J)$ (with $D^{(0)}(\alpha J)=0$) are the rates of elastic collisions that induce transitions between magnetic sublevels belonging to the same $|\alpha J>$ level, $C_I^{(K)}(\alpha J, {\alpha}_l J_l)$ denote the multipolar components of the inelastic collisional rates that induce transitions between a lower level $|{\alpha}_l J_l M_l>$ and the level $|{\alpha} J M>$, and $C_S^{(K)}(\alpha J, {\alpha}_u J_u)$ are the multipolar components of the superelastic collisional rates that induce transitions between the upper level $|{\alpha}_u J_u M_u>$ and the level $|{\alpha} J M>$. 

Given that in this paper we are especially interested in the collisional transfer rates between the $J$-levels of each individual term, we actually have that $\alpha=\alpha_l=\alpha_u$ in Eq. (5). Collisional transitions between the lower and upper levels of the spectral lines under consideration induced by inelastic collisions with electrons are disregarded here (i.e., we assume that the atomic excitation is dominated by radiative transitions). Therefore, here we are investigating a problem where the radiative transitions between the lower and upper levels of the spectral lines under consideration induce population imbalances among the magnetic sublevels of such levels, which are modified by the presence of isotropic collisions with neutral hydrogen atoms, both by the elastic collisions and by the collisional transfer rates between ``nearby" $J$-levels only (i.e., between the $J$-levels belonging to each spectral term of the multilevel model under consideration). 

Note that a different notation is used in the previous papers
by Derouich et al. (e.g., Derouich et al. \cite{Derouich_04}).  There the symbol $D^0(\alpha J \to \alpha J')$ is used to denote the population transfer rates from $J$ to $J'$, while $D^{K}(\alpha J \to \alpha J')$  refers to the polarization transfer rates. The relationship between the two notations is: 
\begin{eqnarray} \label{eq_3}
D^{K}(\alpha J \to \alpha J') & = & \sqrt{ \frac{2J+1} {2J'+1}} \; C^{(K)}(\alpha  J', \alpha  J).  
\end{eqnarray} 

We obtain the $\rho^K_0(J)$ elements corresponding to each $J$-level of the atomic model under consideration by assuming statistical equilibrium (i.e., $\big[\frac{d \; \rho_0^{K} (\alpha J)}{dt}\big]_{rad} + \big[\frac{d \; \rho_0^{K} (\alpha J)}{dt}\big]_{coll} = 0$). Since the ensuing algebraic system of equations in the unknowns $\rho^K_0(J)$ is singular, one of the equations is substituted with that expressing the conservation of particles (i.e., with $\sum_{J}\rho^K_0(J)\sqrt{2J+1}=1$). We have carried out this type of calculations imposing the above-mentioned anisotropic illumination conditions for the following cases: (a) neglecting collisions, (b) considering only the elastic collisions and (c) including also the inelastic and superelastic collisional transfer rates between nearby $J$ levels (i.e., between those pertaining to the same spectral term).

From the computed $\rho^K_0(J)$ values we need to use only the $K=0$ and $K=2$ components corresponding to the upper and lower levels of the chosen spectral line, since these are the density matrix elements needed to compute the $I$ and $Q$ components of the emission vector and of the propagation matrix that appear in the Stokes-vector radiative transfer equation (see Eqs. 7.21 in Landi Degl'Innocenti \& Landolfi 2004). The emissivity contributions, which are proportional to the density matrix elements of the upper level, are given by

\begin{equation}
  \epsilon^{\rm line}_I \,=\,\epsilon_0 [\rho^0_0(J_u)+
	w_{J_uJ_\ell}^{(2)} \frac{1}{2\sqrt{2}} (3\mu^2-1)\rho^2_0(J_u)],\\
\end{equation}
\begin{equation}
  \epsilon^{\rm line}_Q \,=\,\epsilon_0 w_{J_uJ_\ell}^{(2)} \frac{3}{2\sqrt{2}}(1-\mu^2)\rho^2_0(J_u),  
\end{equation}
where $\epsilon_0=(h\nu/4\pi)~A_{u\ell}\sqrt{2J_u+1}\,\,N$ 
($N$ being the number density of the atoms under consideration), and
$w_{J_uJ_\ell}^{(2)}$ is a numerical factor that depends only on the 
quantum numbers of the levels involved in the transition (e.g., see Table 10.1 in Landi Degl'Innocenti \& Landolfi 2004). The absorption and dichroism contributions, which are proportional to the density matrix elements of the lower level, are given by

\begin{equation}
\eta^{\rm line}_I \,=\,\eta_0 [\rho^0_0(J_{\ell})+
w_{J_\ell J_u}^{(2)} \frac{1}{2\sqrt{2}} (3\mu^2-1)\rho^2_0(J_{\ell})],\\
\end{equation}
\begin{equation}
\eta^{\rm line}_Q \,=\,\eta_0 w_{J_\ell J_u}^{(2)} \frac{3}{2\sqrt{2}}(1-\mu^2)\rho^2_0(J_{\ell}),       
\end{equation}
where $\eta_0=(h\nu/4\pi)~B_{\ell u}\sqrt{2J_\ell+1}\;N$. Note that in the expressions for $\epsilon_Q$ and $\eta_Q$ the reference direction for Stokes $Q$ has been chosen as the perpendicular to the radial direction through the observed point.

As in Manso Sainz \& Landi Degl'Innocenti (2002), we calculate the emergent fractional linear polarization in the limit of tangential observation in a plane-parallel atmosphere ($90^{\circ}$ scattering geometry). For such a line-of-sight the solution of the transfer equations for Stokes $I$ and $Q$ is

\begin{equation}
   \left(\frac{Q}{I}\right)_{\mu\rightarrow 0}\,=\,
	\frac{\epsilon_Q/\epsilon_I-\eta_Q/\eta_I}
	{1-(\eta_Q/\eta_I)(\epsilon_Q/\epsilon_I)}\, ,    
\end{equation}
where $\epsilon_X$ and $\eta_X$ (with $X=I$ or $X=Q$)
must be evaluated at the surface point of the
plane-parallel slab. Actually, given that the solar atmosphere is a weakly anisotropic medium, this expression can be simplified as follows:

\begin{equation}
   \left(\frac{Q}{I}\right)_{\mu\rightarrow 0}\,{\approx}\,\,
	{\epsilon_Q/\epsilon_I-\eta_Q/\eta_I},
\end{equation}	
which shows clearly that the observed scattering polarization in a given spectral line has, in general, two contributions (Trujillo Bueno 1999): one due to the selective emission of polarization components resulting from the polarization of the upper level, and an extra one caused by dichroism --that is, by the selective absorption of polarization components resulting from the lower level polarization.

Obviously, when the lower level of the spectral line under consideration is unpolarized, then the only contribution to the 
emergent fractional linear polarization would be that caused by the selective emission of polarization components that result from the population imbalances of the upper-level. Under such circumstances

\begin{equation}
   \left(\frac{Q}{I}\right)_{\mu\rightarrow 0}\,{\approx}\,\,
	{\epsilon_Q/\epsilon_I}\,{\approx}\,\frac{3}{2\sqrt{2}}\,w_{J_uJ_\ell}^{(2)}\,(1-\mu^2)\,[\rho^2_0(J_u)/\rho^0_0(J_u)],
\end{equation}	
where the approximate expression applies to the line center 
of a sufficiently strong spectral line.
In Fig. 2 below we will show what happens when we make this unpolarized lower-level assumption. In addition, we will also show what happens when the lower level is allowed to be polarized. To this end, we only need to solve the statistical equilibrium equations without introducing any collisional depolarization. An important point to remember here is that the presence of lower level atomic polarization may have a significant feedback on the polarization of the upper levels, so that a calculation of the emergent $Q/I$ neglecting
the dichroism contribution (i.e., ignoring the $\eta_Q/\eta_I$ term of Eq. 12) will give in general a $Q/I$ amplitude different to that corresponding to the unpolarized lower level case (Trujillo Bueno 1999, 2001). 

\section{Application to multiplet 42 of Ti {\sc i}}

This section describes the results of our investigation of the influence of elastic, inelastic and superelastic collisions on the atomic polarization of the lower and upper levels of the 13 lines of multiplet 42 of Ti {\sc i}, and on the amplitudes of the ensuing emergent fractional linear polarization. We begin by considering the sensitivity of the calculated fractional linear polarization to the complexity of the assumed multilevel model, justifying why we end up using a 10 level model with the five $J$ levels of the lower term $^3F$ and the five $J$ levels of the upper term $^3F^0$ of multiplet 42 of Ti {\sc i}. We then investigate the sensitivity of the emergent linear polarization to the above-mentioned collisions. Appendix A gives polynomial approximants for calculating the collisional rates for any combination of hydrogen number density and temperature of the solar photospheric plasma. Such collisional rates, which are given here in ${\rm s}^{-1}$, result from the application of the semi-classical theory used by Derouich et al. (2005b).


We have considered two atomic models. The most realistic one (hereafter the full model) is similar to that used by Manso Sainz \& Landi Degl'Innocenti (2002), which has 395 $J$ levels and all the allowed transitions between them. The total number of $\rho^K_0(J)$ elements necessary to describe the excitation state of the full atomic model is 1564, whose values can be found by solving the above-mentioned statistical equilibrium equations for given illumination conditions (i.e., using the $J^0_0$ and $J^2_0$ values computed from the center-to-limb variation of the solar continuum radiation tabulated by Cox (2000) for each allowed radiative transition)\footnote{Figure 2 of Manso Sainz \& Landi Degl'Innocenti (2002) shows the variation with wavelength of the anisotropy factor, $w=\sqrt{2}J^2_0/J^0_0$, and of the number of photons per mode (i.e., $\bar{n}=J^0_0(c^2/2h\nu^3)$) of the solar continuum radiation.}. In any case, given the low degree of radiation anisotropy in the solar atmos
 phere, one can safely solve the statistical equilibrium equations neglecting the $\rho^K_0(J)$ elements with $K>2$ (e.g., Manso Sainz \& Landi Degl'Innocenti 2002), thus reducing the number of $\rho^K_0(J)$ unknowns to 771. For this reason, in the appendices we have given the collisional rates only for $K=0$ and $K=2$.

The second multilevel model we have used is that shown in Fig. 1, which contains all the lower and upper $J$ levels of multiplet 42 of Ti {\sc i}. The number of $\rho^K_0(J)$ elements necessary to describe the excitation state of this 10 level model atom is 20 when those with $K>2$ are neglected. It is important to note that the lower levels of the 13 lines of this multiplet are metastable, so that for a given collisional rate the population imbalances of such long-lived levels are much more vulnerable to collisions than the short-lived upper levels of the spectral lines under consideration. 

\begin{figure}
\begin{center}
\includegraphics[width=8 cm]{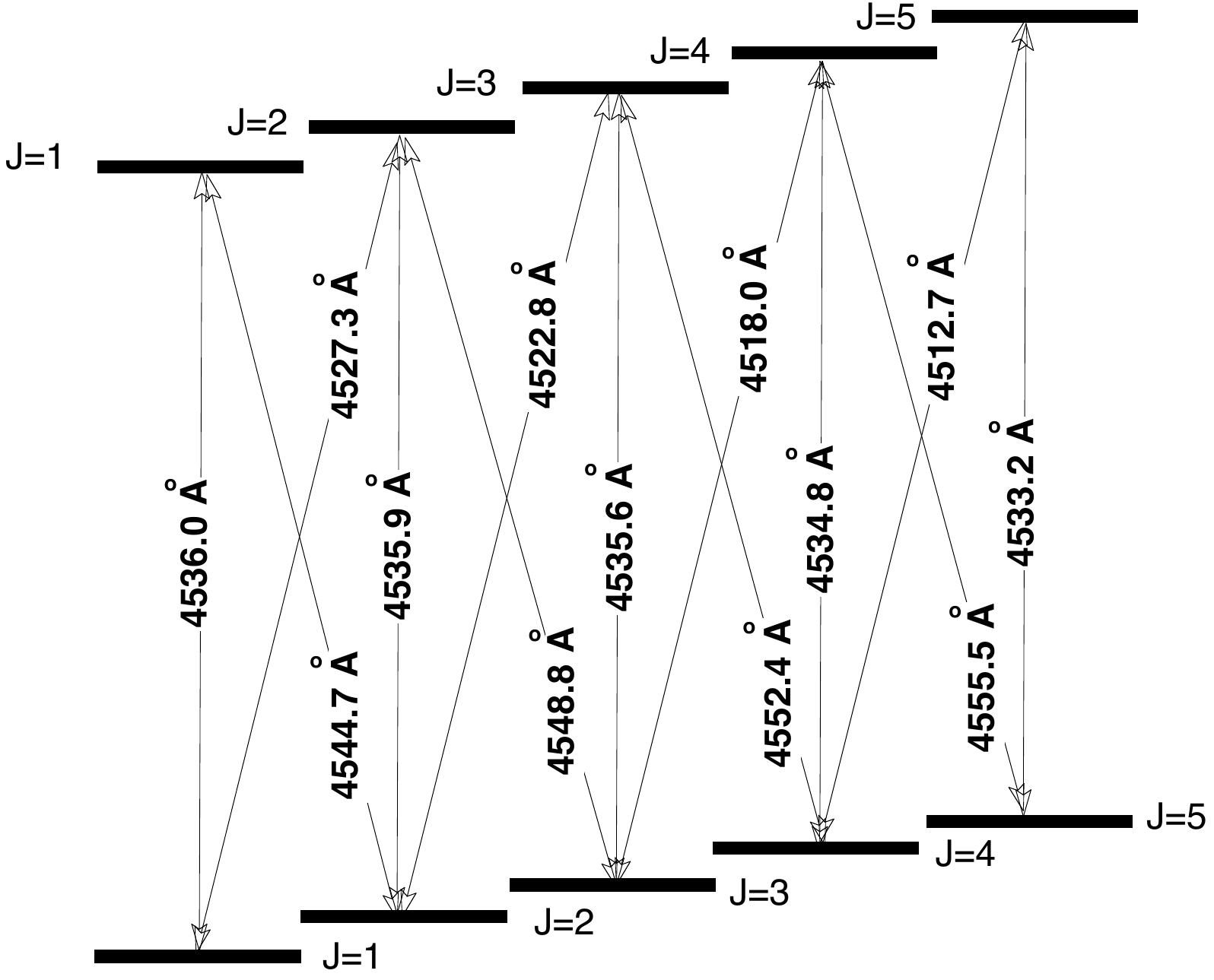}
\end{center}
\caption{Grotrian diagram of multiplet 42 of \ion{Ti}{i} showing the angular momentum values of the levels of the lower term $a \; ^5F$ and of those of the upper term $y \, ^5F^o$. The wavelengths (in Angstroms units) of the 13 lines of this multiplet are also indicated.}
\label{figure1}
\end{figure}

Figure 2 shows the results of calculations with the two multilevel models mentioned above, neglecting the collisional rates (i.e., taking into account only the radiative transitions). The upper panel neglects the dichroism contribution caused by the selective absorption of polarization components (i.e., the emergent polarization amplitudes were obtained by using Eq. 13, which neglects the $\eta_Q/\eta_i$ term of Eq. 12). The lower panel was obtained using instead Eq. (13), so that it takes into account that selective absorption contribution, in addition to the usual one caused by the selective emission of polarization components resulting from the population imbalances of the upper level. As seen in Fig. 2 the 10 level model atom significantly overestimates the fractional linear polarization of the ${\lambda}4536$ line, when the role of collisions is neglected. As pointed out by Manso Sainz et al. (2004) this Ti {\sc i} line is insensitive to the presence of magnetic fields b
 ecause the Land\'e factors of its upper and lower levels are exactly zero. Therefore, it is a very useful reference line whose polarization pattern can be used to investigate the reliability of the thermal and dynamic properties of solar photospheric models (see Shchukina \& Trujillo Bueno 2007).

\begin{figure}
\begin{center}
\includegraphics[width=8 cm]{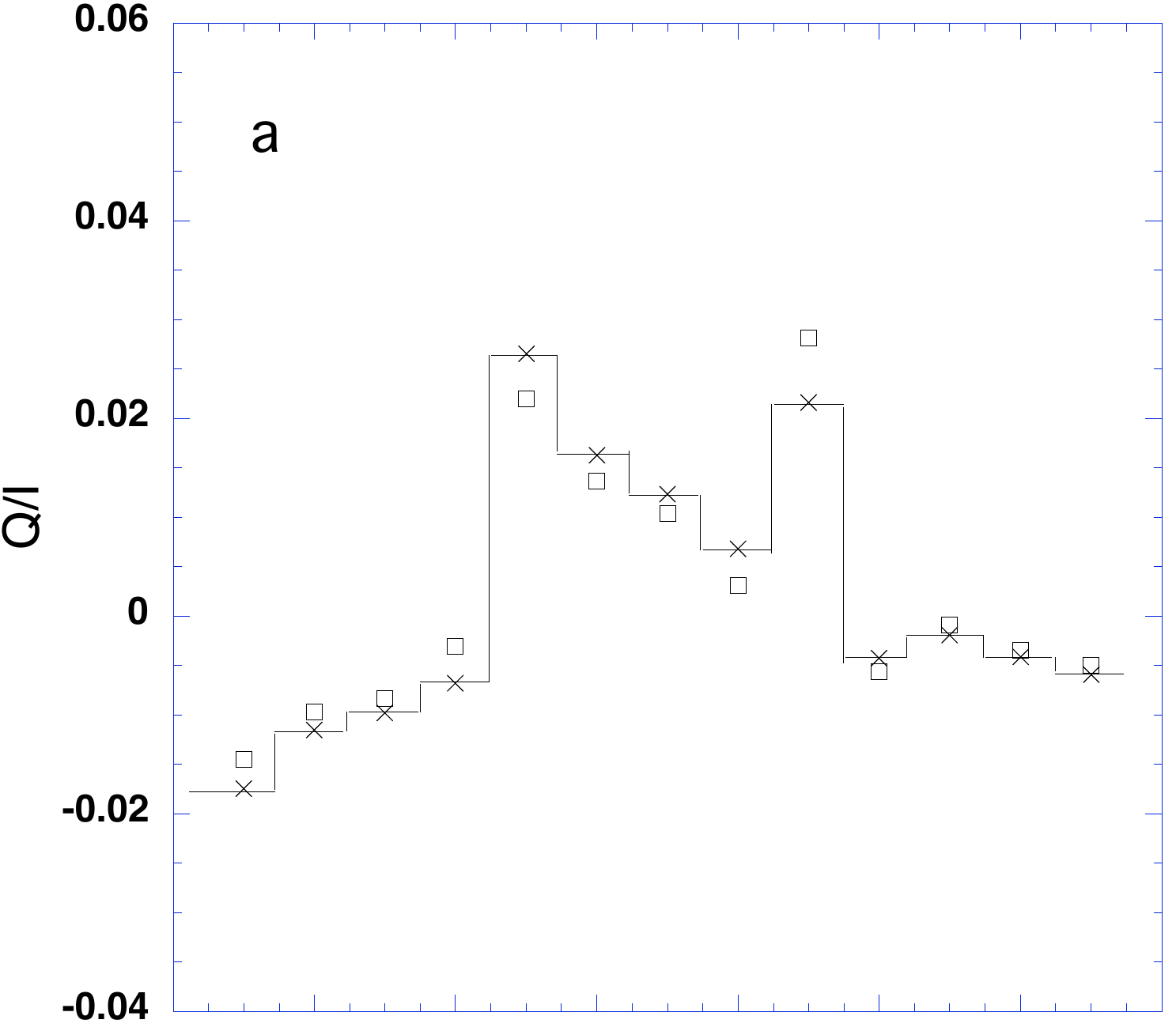}
\includegraphics[width=8 cm]{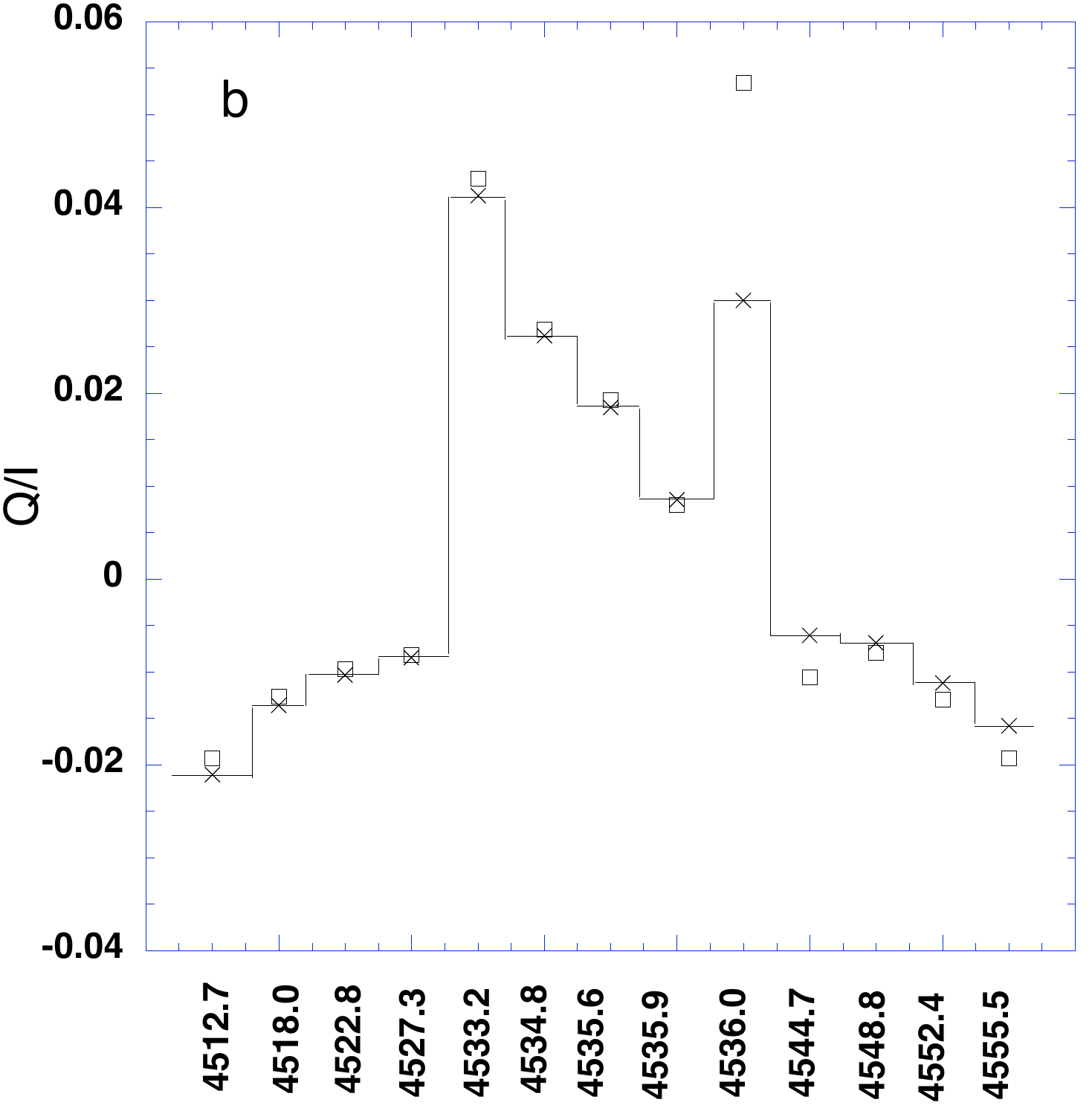}
\end{center}
\caption{Emergent fractional linear polarization amplitudes, $Q/I$, for each of the 13 lines of multiplet 42 of Ti {\sc i} indicated in Fig. 1. $\square$: 10 levels model.  $\times$: full multilevel model. Collisions are not taken into account. Panel (a) shows the emergent fractional polarizations if dichroism is neglected (i.e., using Eq. 13). Panel (b) shows the emergent fractional polarizations if dichroism is taken into account, in addition to the standard contribution of selective emission of polarization components (i.e., using Eq. 12). Note that in panel (b) the polarization amplitudes corresponding to the calculation performed using the full multilevel model agree with those shown by Manso Sainz et al. (2004) in their Fig. 2b.}
\label{figure2}
\end{figure}

As shown in Fig. 3, when the lower metastable levels are assumed to be completely unpolarized then the 10 level model atom gives excellent results. Interestingly, as pointed out by Shchukina \& Trujillo Bueno (2007), this seems to be indeed the case in the solar atmosphere. In fact, the elastic collisional rates given in Appendix A for the 5 lower levels of multiplet 42 of Ti {\sc i} are such that $\delta(J_l)=D^{(2)}(J_l)\,t_{\rm life}(J_l)\,{\gg}\,1$ for each lower-level $J_l$, even when using the relatively low hydrogen number densities of the temperature minimum region of standard semi-empirical atmospheric models (i.e., $n_H\,{\approx}\,2{\times}10^{15}\,{\rm cm}^{-3}$). For example, $1/t_{\rm life}(J_l)\,{\sim}\,B_{lu}J^0_0\,{\sim}\,B_{lu}B_{\nu}(T)/2\,{\sim}\,1.7{\times}10^5\,{\rm s}^{-1}$ (when using 
$\lambda=4536\AA$ and $T=5800$ K). On the other hand, $D^{(2)}(J_l)\,{\sim}\,8{\times}10^{-9}\,n_H$ (see Appendix A). Therefore, $\delta(J_l)\,{\sim}\,1$ for $n_H\,{\approx}\,2{\times}10^{13}\,{\rm cm}^{-3}$, while $\delta(J_l)\,{\sim}\,10^2$ in the denser solar atmospheric region where the line-center polarization of such Ti {\sc i} lines originates. Accordingly, from now on we will use the above-mentioned 10 level model atom without lower-level polarization
in order to investigate the impact of collisions on the population imbalances of the 5 upper levels of Fig. 1, and on the emergent linear polarization in the ensuing 13 lines. Finally, it is also of interest to note the differences between the polarization amplitudes shown in Figs. 2a, 2b and 3. In order to understand why they are different, it suffices to take into account what we have mentioned in the last paragraph of Section 2.

\begin{figure}
\begin{center}
\includegraphics[width=8 cm]{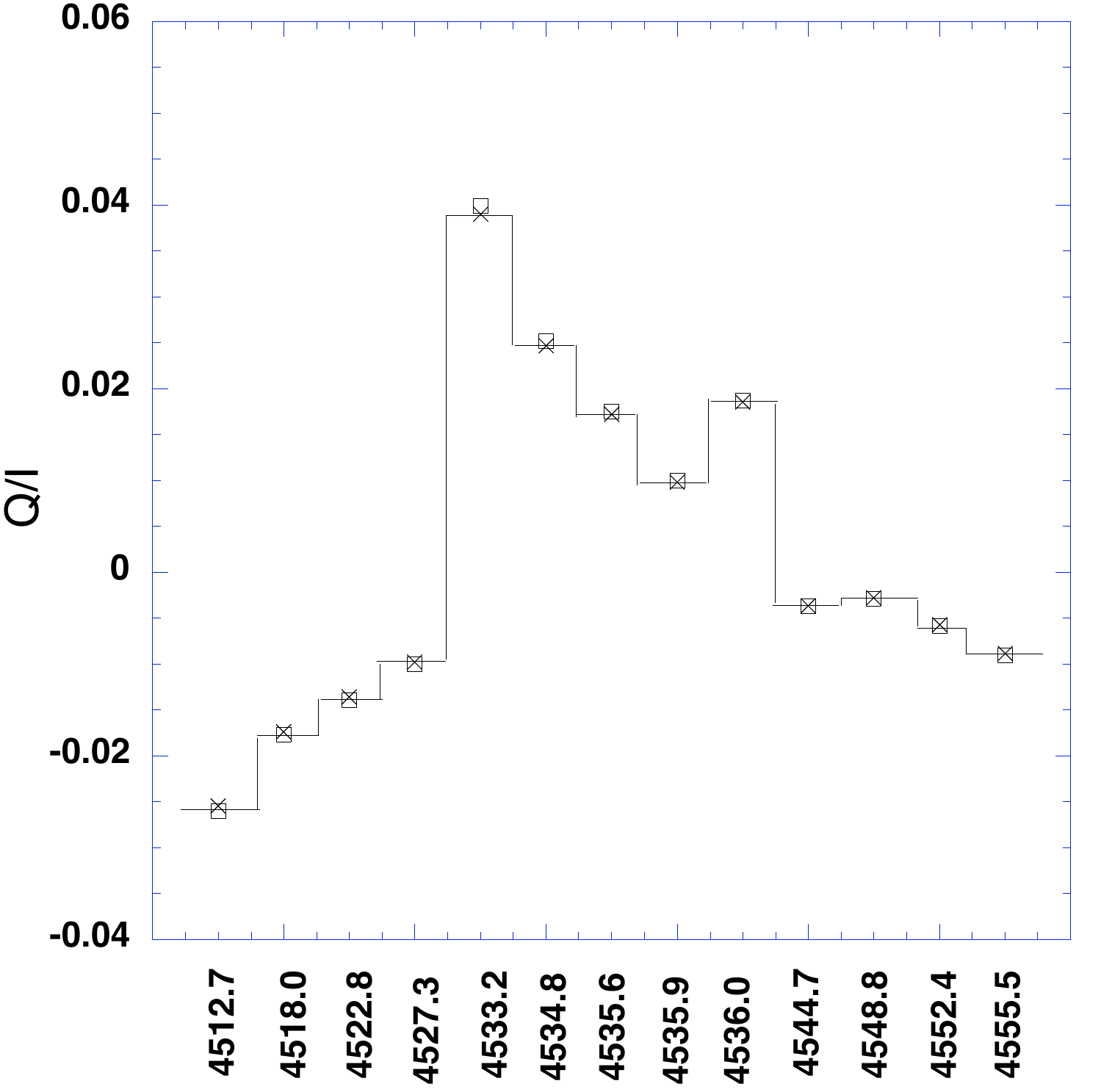}
\end{center}
\caption{Emergent fractional linear polarization amplitudes, $Q/I$, for each of the 13 lines of multiplet 42 of Ti {\sc i} indicated in Fig. 1.
$\square$: 10 levels model.  $\times$:  full atomic model. The atomic polarization of all (long-lived) lower levels is assumed to be zero. The influence of collisions on the population imbalances of the upper levels is not taken into account.}
\label{figure3}
\end{figure}


In order to investigate the impact of the collisional rates 
on the emergent fractional linear polarization amplitudes
we have done the following. For each spectral line we use the hydrogen number density and the kinetic temperature values corresponding to the height $h$ in the quiet Sun photospheric reference model of Maltby et al. (1986) where the line-center optical depth is unity for a simulated observation along a line-of-sight specified by $\mu=0.1$ (with $\mu$ the cosine of the heliocentric angle). Such heights $h$ for each line of multiplet 42 of Ti {\sc i} are given in Fig. 5 of Trujillo Bueno et al. (2006), where it is shown that they are contained in the range between 400 and 550 km. For example, $h{\approx}450$ km for the ${\lambda}4536$ line (Shchukina \& Trujillo Bueno 2007). 

Figure 4 shows the calculated fractional polarization amplitudes in the presence of collisions. Elastic collisions with neutral hydrogen atoms (i.e., the $D^{(2)}(J)$ rates) have a negligible influence on the upper levels of all the lines of the titanium multiplet under consideration, while the collisional transfer rates have a small but noticeable depolarizing impact on a few of them (i.e., mainly on the upper levels with $J_u=1$ and $J_u=2$). This is because the population transfer rate between the  $J_u=1$ and $J_u=2$ levels is substantially larger than the rest of the collisional transfer rates (see Appendix A.4). As seen in Fig. 4, the observable effect of such an atomic level depolarization is mainly seen in the ${\lambda}4536$ line (i.e., that of Fig. 1 with $J_u=J_l=1$), in spite of the fact that the ${\lambda}4544.7$ line (i.e., that with $J_u=1$ and $J_l=2$) has the same upper level. 
For both lines we find

\begin{eqnarray}  
\frac{ {(Q/I)}_\textrm{all collisional rates} } { {(Q/I)}_\textrm{only the collisional elastic rates} }= 0.75,
\end{eqnarray} 
but in Fig. 4 this is only noticed in the $Q/I$ of the ${\lambda}4536$ line. The reason for this behaviour can be easily understood through Eq. (13), which implies that $\Delta(Q/I){\approx}(3/2\sqrt{2})w_{J_uJ_l}^{(2)}\Delta[{\rho^2_0(J_u)/\rho^0_0(J_u)}]$, with $w_{11}^{(2)}=-0.5$ for the ${\lambda}4536$ line and $w_{12}^{(2)}=0.1$ for the ${\lambda}4544.7$ line. Our conclusion is that the collisional transfer rates among the upper $J$ levels of multiplet 42 of Ti {\sc i} have a small but noticeable depolarizing impact on the emergent fractional linear polarization of the ${\lambda}4536$ line. As mentioned above, this is a truly remarkable spectral line because its polarization turns out to be virtually insensitive to the presence of magnetic fields (see Fig. 2 of Manso Sainz et al. 2004). We have checked that the alignment transfer rates among the upper levels of the multiplet do not change this conclusion (see, e.g., Fig. 4). 

\begin{figure}
\begin{center}
\includegraphics[width=8 cm]{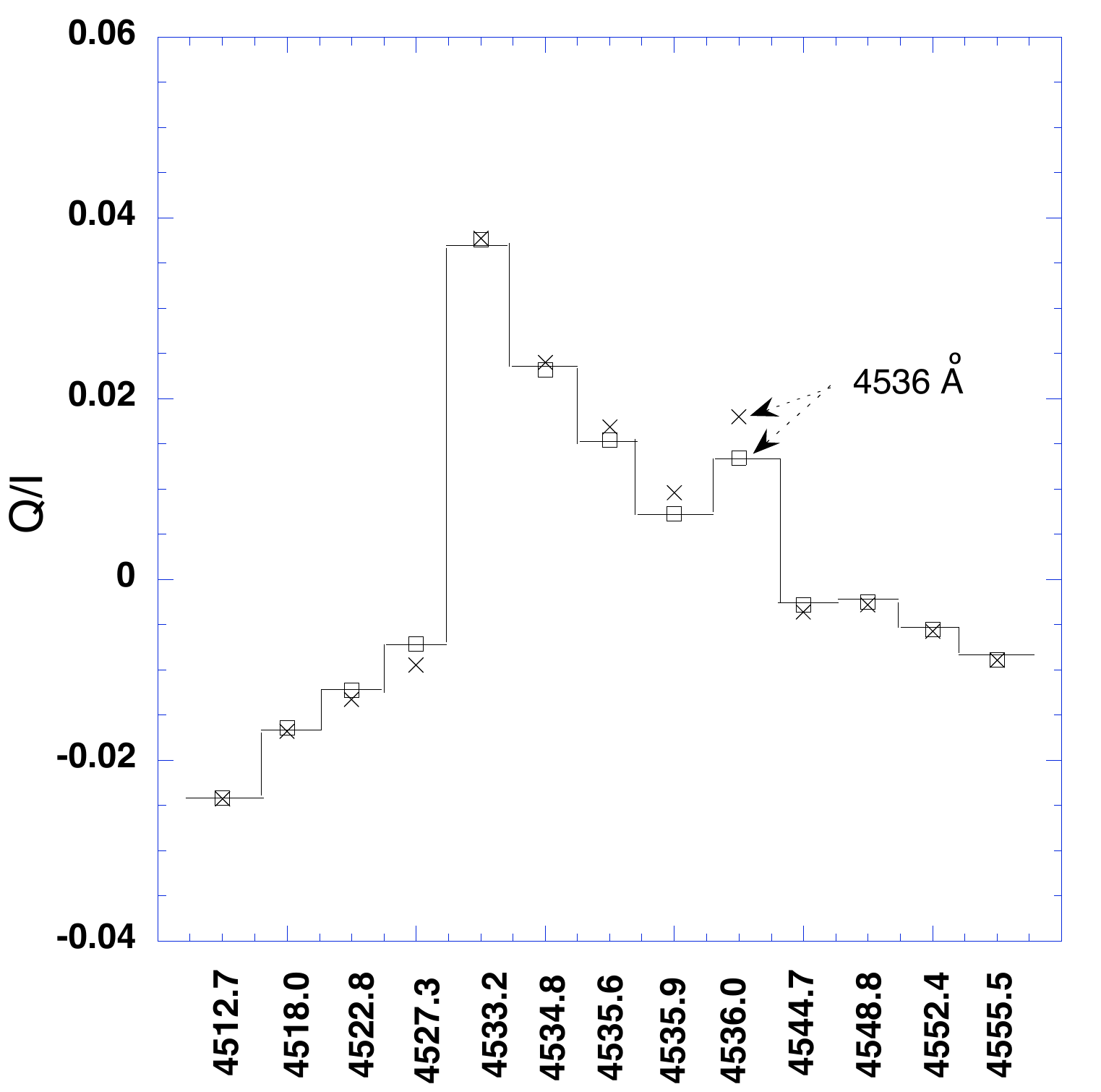}
\end{center}
\caption{Emergent fractional linear polarization amplitudes, $Q/I$, for each of the 13 lines of multiplet 42 of Ti {\sc i} using the 10 levels model atom shown in Fig. 1 without lower-level polarization. $\times$: $Q/I$ amplitudes taking into account only $D^{(2)}(J) $. $\square$: $Q/I$ values using all collisional rates: $D^{(2)}(J)$, $C^{(0)}(\alpha J',  \alpha J)$ and  $C^{(2)}(\alpha J,  \alpha J')$.}
\label{figure5}
\end{figure}

\section{Application to Ca {\sc ii} }

This section describes the results of our investigation of the influence of collisions on the atomic polarization of the levels of the Ca {\sc ii} model of Fig. 5. In addition to the elastic collisions, we also take into account the collisional transfer rates between the two $J$ levels of the term $^2P$ (which are the upper levels of the H and K lines)
and between those belonging to the term $^2D$ (which are the lower
levels of the Ca {\sc ii} IR triplet). In contrast with the lower levels of multiplet 42 of Ti {\sc i} the two metastable levels of the Ca {\sc ii} IR triplet (i.e., the $^2D$ levels with $J=3/2$ and $J=5/2$) are polarized. This was demonstrated by Manso Sainz \& Trujillo Bueno (2003) who solved the full radiative transfer problem for the multilevel model atom of Fig. 5 in a semi-empirical model of the solar atmosphere, and taking into account elastic, inelastic and superelastic collisions with the ensuing rates calculated following Shine \& Linsky (1974) for the inelastic and superelastic rates and Lamb \& ter Haar (1971) for the elastic rates. Appendix B gives the collisional rates obtained by Derouich et al. (2004), but using instead the notation adopted in the present paper. Such rates were obtained via the application of the semi-classical theory of collisions.

\begin{figure}
\begin{center}
\includegraphics[width=8 cm]{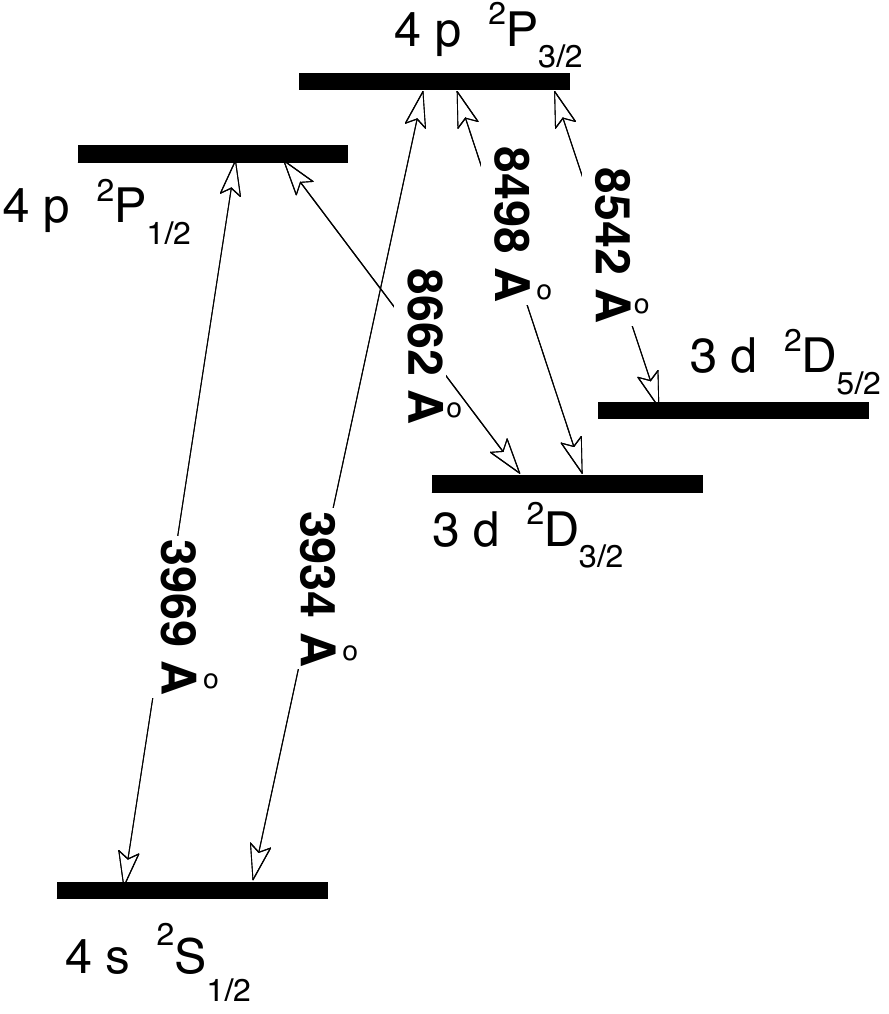}
\end{center}
\caption{Grotrian diagram of \ion{Ca}{ii} showing the upper and lower $J$-levels of the H and K lines and of the IR triplet.}
\label{figure6}
\end{figure}

\begin{table}
\caption{Some of the values used in the Ca {\sc ii} calculations}
\begin{center}
\begin{tabular}{cccccc}
\hline
\hline
line & H    &    K    &  8498 &   8542 & 8662 \\
\hline
$h$ (Km) &      2040 & 2000 & 1200 & 1460 & 1380    \\
$n_H$ (cm$^{-3}$) &  $1.4 \times 10^{11}$  &  1.7 $10^{11}$& $7.6 \times 10^{12}$ & $1.7 \times 10^{12}$  &$2.5 \times 10^{12}$ \\
$T$(K)   &       8780 & 8550 & 6475 & 6930 & 6800   \\
$\bar{n} $&  $0.12   \times 10^{-3}$     &  $0.13  \times 10^{-3}$&  $13 \times 10^{-3}$  & $8.7 \times 10^{-3}$   &    $9.1\times 10^{-3}$ \\
$w$ & 0.031     & 0.031  & 0.012  & 0.028  &  0.036 \\
\hline
\end{tabular}
\end{center}
\end{table}


In order to investigate the impact of collisions on the Ca {\sc ii} lines we have applied the following strategy, which relies on the results that Manso Sainz \& Trujillo Bueno (2003) obtained through detailed multilevel radiative transfer calculations in a 
semi-empirical atmospheric model similar to that of Fontenla et al. (1993). For each spectral line indicated in Fig. 5
Table 1 gives the total hydrogen number density, the kinetic temperature, ${\bar{n}}$ (i.e., the number of photons per mode of the radiation field) and $w=\sqrt{2}J^2_0/J^0_0$ (i.e., the anisotropy factor) at the atmospheric height in the plane-parallel semi-empirical model where the line center optical depth is unity for a simulated observation at $\mu=0.1$. We performed four multilevel calculations in order to compute the emergent $Q/I$ in each of the four spectral lines that can in principle be polarized\footnote{Note that there can be no scattering polarization in the H line because its lower and upper levels cannot be aligned.}. For example, for the ${\lambda}8542$ line we solved the statistical equilibrium equations for the 5-level model atom of Fig. 5 using the $n_H$, $T$, ${\bar{n}}$ and $w$ values corresponding to the height $h=1460$ Km. These values at $h=1460$ Km are given in Table 1 for the ${\lambda}8542$ line, but we have used also the ${\bar{n}}$ and $w$ values
  corresponding to each of the remaining line transitions at the very same height. The resulting $\rho^2_0$ values corresponding to the lower and upper levels of the ${\lambda}8542$ line were then used to compute $Q/I$ according to Eq. (12). A similar calculation was performed for each of the other three lines that can be polarized by scattering processes (i.e., the K line, the 8498 \AA\ line or the 8662 \AA\ line), but solving the statistical equilibrium equations using for each of the line transitions of Fig. 5 the $n_H$, $T$, ${\bar{n}}$ and $w$ values corresponding to the atmospheric height $h$ where the line center optical depth along $\mu=0.1$ is unity for the particular spectral line under consideration.

The results are shown in Fig. 6 for the collisionless case, for the case in which only elastic collisions are taken into account, and for the most general situation in which also the collisional transfer rates are considered. As seen in the figure, the collisional transfer rates play only a minor role on the scattering polarization of the Ca {\sc ii} IR triplet, while they are completely insignificant for the K line at 3934 \AA.  

\begin{figure}
\begin{center}
\includegraphics[width=8 cm]{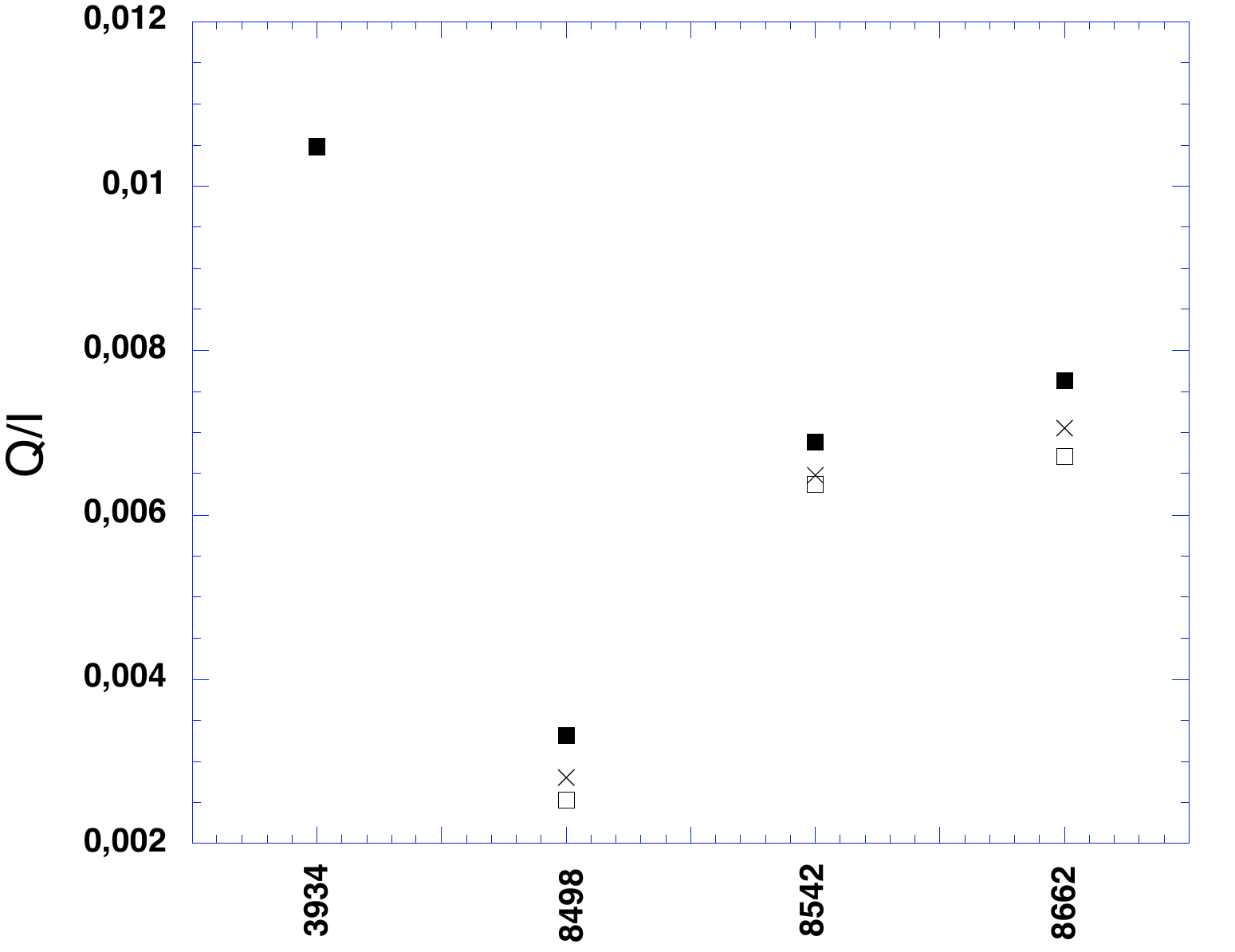}
\end{center}
\caption{Emergent fractional linear polarization amplitudes, $Q/I$, for the K line and for the IR triplet of Ca {\sc ii}. The calculations have been performed using the multilevel model of Fig. 5, and following the strategy indicated in the text. $\blacksquare$: $Q/I$ values in the absence of collisions.  $\times$: $Q/I$ amplitudes taking into account only $D^{(2)}(J)$.  $\square$: $Q/I$ values calculated using all collisional rates: $D^{(2)}(J)$, $C^{(0)}(\alpha J',  \alpha J)$ and $C^{(2)}(\alpha J,  \alpha J')$.}
\label{figure7}
\end{figure}


Finally, we would like to comment briefly 
on an interesting point about the K-line polarization.
It is commonly believed that isotropic collisions have always a depolarizing role. However, when the lower levels of the line transitions under consideration are polarized, we may find circumstances where isotropic collisions increase the emergent polarization amplitudes. This was first pointed out by Trujillo Bueno \& Landi Degl'Innocenti (1997) for the two-level atom case with lower-level polarization, but other authors have supported this conclusion through their calculations for more realistic model atoms (e.g., Manso Sainz \& Landi Degl'Innocenti 2002; Klement \& Stenflo 2003). 

Here we show that there is a range of values of the hydrogen number density for which we have collisional enhancement of the emergent fractional linear polarization in the K line of Ca {\sc ii}. As seen in Fig. 7, the $n_H$ values for which this is possible are larger than those encountered in the region of the solar chromosphere where the K line-core polarization originates, so that collisions with neutral hydrogen atoms can indeed be neglected when modeling the K-line polarization.
In any case, it is important to keep in mind that this possibility exists for some spectral lines of the solar spectrum. 

\begin{figure}
\begin{center}
\includegraphics[height=8.0cm]{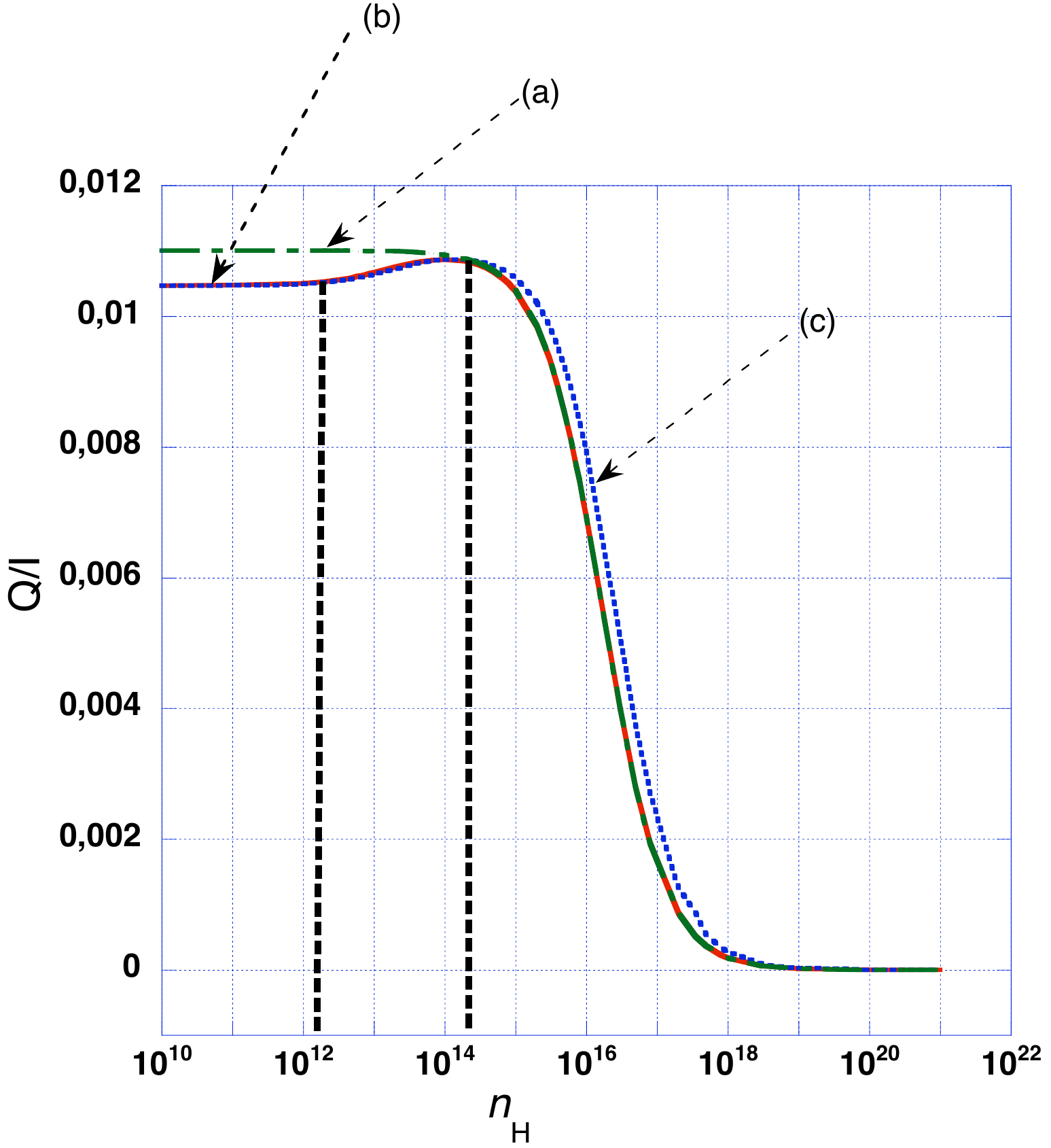}
\end{center}
\caption{Emergent fractional linear polarization amplitude
of the \ion{Ca}{ii} K line as a function of the neutral hydrogen density. (a) Dot-dashed line: 
$Q/I$ values calculated assuming that the metastable levels $^2D_{3/2}$ and $^2D_{5/2}$ are completely unpolarized.
(b) Solid line: $Q/I$ amplitudes calculated using all collisional rates, taking into account the polarization ofsuch metastable levels: $D^{(2)}(J)$, $C^{(0)}(\alpha J',  \alpha J)$ and   $C^{(2)}(\alpha J,  \alpha J')$. 
(c)  Dotted line: same as in (b) but taking into account only $D^{(2)}  (J)$. The dashed vertical lines, between $n_H \simeq 2 \times 10^{12}$ cm$^{-3}$ and $n_H \simeq 2 \times 10^{14}$ cm$^{-3}$, represent  the collisional regime where the scattering polarization in the K line increases with the hydrogen number density. }
\label{figure8}
\end{figure}
 
\section{Concluding comments}

The collisional rates given in the appendices resulted from the application of the semi-classical theory of collisions. From comparisons with the published results obtained through quantum chemistry methods we estimate that any possible error in our collisional rates for Ca {\sc ii} is smaller than $10\%$. Unfortunately, we cannot give a similar error estimate for our collisional rates of the complex Ti {\sc i} atom. The only thing we can say at this stage is that the semi-classical theory of collisions applied to levels where we have just one single valence electron is expected to give fairly accurate results for the intermediate interatomic distances, which are the ones playing the key role in the calculation of the collisional rates (see Fig. 6 of Derouich et al. 2003).

We confirm that elastic collisions with neutral hydrogen atoms in the solar atmosphere destroy the atomic polarization of the (metastable) lower levels of multiplet 42 of Ti {\sc i}. This is important because without lower-level polarization a 10-level model atom is sufficient for modeling the scattering polarization of the 13 lines of such a multiplet via radiative transfer calculations in three-dimensional (3D) models of the solar atmosphere, as done by Shchukina \& Trujillo Bueno (2007). On the other hand, elastic collisions have a rather small depolarizing effect on the (metastable) lower levels of the Ca {\sc ii} IR triplet and a negligible impact on the atomic polarization of the upper levels of all the above-mentioned spectral lines. 

We have also shown that the collisional transfer rates have an  unnoticeable impact on the emergent fractional linear polarization for most of the lines of multiplet 42 of Ti {\sc i}, except perhaps for the ${\lambda}4536$ line of Ti {\sc i} for which it is easier to detect a
$25\%$ reduction in the emergent polarization amplitude. This is important because such a line is insensitive to the presence of magnetic fields, which implies that we can use its scattering polarization pattern for investigating the reliability of 3D (magneto)-hydrodynamical models of the solar photosphere.

Collisional transfer rates might be truly important for other atomic or molecular systems with closer $J$-levels (e.g., hyperfined structured multiplets and/or molecules). In fact, there are both theoretical and empirical indications that they could be significant for molecules like SiO (Derouich 2006) and MgH (Asensio Ramos \& Trujillo Bueno 2005; Trujillo Bueno et al. 2006). Therefore, future theoretical and experimental research in this direction will be worthwhile.


\begin{acknowledgements}

We thank Nataliya Shchukina for valuable discussions. 
We would also like to thank Andr\'es Asensio Ramos for helping us to carry out in a more efficient way the multilevel calculations presented in this paper. Partial support by the Spanish Ministerio de Educaci\'on y Ciencia through project AYA2004-05792 and by the European Solar Magnetism Network is gratefully acknowledged. 
 
\end{acknowledgements}


\begin{appendix}   

\section{Multiplet 42 of neutral Titanium}  

In what follows only the elastic and inelastic collisional rates (in ${\rm s}^{-1}$) are given as a function of neutral hydrogen number density $n_H$ (in ${\rm cm}^{-3}$) and kinetic temperature $T$ (in Kelvins).
However, it is straightforward to retrieve the values of the superelastic collisional rates $C_S^{(K)}(\alpha_l J_l,\alpha_u J_u)$ by applying  the relation:
\begin{eqnarray} 
C_S^{(K)}(\alpha_l J_l,\alpha_u J_u)= \frac{2J_l+1}{2J_u+1} \exp\frac{E_{\alpha_u J_u}-E_{\alpha_l J_l}}{k_B T}  \; C_I^{(K)}(\alpha_u J_u,\alpha_l J_l),  \nonumber  
\end{eqnarray} 
$E_{\alpha J}$ being the energy of the level $(\alpha J)$ and  $k_B$ is the Boltzmann constant.

To simplify the notation we omit writting $\alpha$ in the collisional rate
symbol. Note that $C_I^{(K)}(J, J_l)$ denotes the multipolar component of the inelastic collisional rates that induce transitions between a lower level $|J_l M_l>$ and the level $|J M>$ (both belonging to the same spectral term). Note that the largest population transfer rate is obtained for the lowest $J_l$ value and that
the semi-classical theory gives zero collisional transfer rates for $\Delta{J}=J-J_l{>}2$. With obvious notation, these are the collisional rates for the Ti {\sc i} levels of Fig. 1:

\subsection{Elastic collisional rates: lower  levels}

\begin{eqnarray} 
D^{(2)}(1) &=& 8.69 \times 10^{-9} \; n_H (\frac{T}{5000})^{0.42}  \;\textrm{s}^{-1} \nonumber \\
D^{(2)}(2) &=& 8.35 \times 10^{-9} \; n_H (\frac{T}{5000})^{0.42}  \;\textrm{s}^{-1} \nonumber \\
D^{(2)}(3) &=& 8.19 \times 10^{-9} \; n_H (\frac{T}{5000})^{0.41}  \;\textrm{s}^{-1} \nonumber \\
D^{(2)}(4) &=& 7.35 \times 10^{-9} \; n_H (\frac{T}{5000})^{0.42}  \;\textrm{s}^{-1} \nonumber \\
D^{(2)}(5) &=& 7.46 \times 10^{-9} \; n_H (\frac{T}{5000})^{0.41}  \;\textrm{s}^{-1} \nonumber
\end{eqnarray} 

\subsection{Population and alignment transfer rates: lower  levels}

\begin{eqnarray} 
 C_I^{(0)}(2, 1) &=& 11.04 \times 10^{-9} \; n_H (\frac{T}{5000})^{0.42}  \;\textrm{s}^{-1} \nonumber \\
C_I^{(2)}(2, 1) &=& 3.22 \times 10^{-9} \; n_H (\frac{T}{5000})^{0.44}  \;\textrm{s}^{-1} \nonumber \\
C_I^{(0)}(3,1) &=& 5.31 \times 10^{-9} \; n_H (\frac{T}{5000})^{0.41}  \;\textrm{s}^{-1} \nonumber \\
C_I^{(2)}(3, 1) &=& 4.37 \times 10^{-9} \; n_H (\frac{T}{5000})^{0.47}  \;\textrm{s}^{-1} \nonumber \\
C_I^{(0)}(3, 2) &=& 3.05 \times 10^{-9} \; n_H (\frac{T}{5000})^{0.40}  \;\textrm{s}^{-1} \nonumber \\
C_I^{(2)}(3, 2) &=& 2.01 \times 10^{-9} \; n_H (\frac{T}{5000})^{0.46}  \;\textrm{s}^{-1} \nonumber \\
C_I^{(0)}(4, 2) &=& 2.91 \times 10^{-9} \; n_H (\frac{T}{5000})^{0.42}  \;\textrm{s}^{-1} \nonumber \\
C_I^{(2)}(4, 2) &=& 2.77 \times 10^{-9} \; n_H (\frac{T}{5000})^{0.45}  \;\textrm{s}^{-1} \nonumber \\
C_I^{(0)}(4, 3) &=& 3.02 \times 10^{-9} \; n_H (\frac{T}{5000})^{0.42}  \;\textrm{s}^{-1} \nonumber \\
C_I^{(2)}(4, 3) &=& 2.99 \times 10^{-9} \; n_H (\frac{T}{5000})^{0.43}  \;\textrm{s}^{-1} \nonumber \\
C_I^{(0)}(5, 3) &=& 2.95 \times 10^{-9} \; n_H (\frac{T}{5000})^{0.42}  \;\textrm{s}^{-1} \nonumber \\
C_I^{(2)}(5, 3) &=& 2.88 \times 10^{-9} \; n_H (\frac{T}{5000})^{0.44}  \;\textrm{s}^{-1} \nonumber \\
C_I^{(0)}(5, 4) &=& 3.33 \times 10^{-9} \; n_H (\frac{T}{5000})^{0.42}  \;\textrm{s}^{-1} \nonumber \\
C_I^{(2)}(5, 4) &=& 3.12 \times 10^{-9} \; n_H (\frac{T}{5000})^{0.41}  \;\textrm{s}^{-1} \nonumber 
\end{eqnarray}
 
\subsection{Elastic collisional rates: upper levels}
\begin{eqnarray}
 D^{(2)}(1) &=& 1.21 \times 10^{-9} \; n_H (\frac{T}{5000})^{0.44} \;\textrm{s}^{-1} \nonumber \\
D^{(2)}(2) &=& 1.00 \times 10^{-9} \; n_H (\frac{T}{5000})^{0.45} \;\textrm{s}^{-1} \nonumber \\
D^{(2)}(3) &=& 0.81 \times 10^{-9} \; n_H (\frac{T}{5000})^{0.47}  \;\textrm{s}^{-1} \nonumber \\
D^{(2)}(4) &=& 0.93 \times 10^{-9} \; n_H (\frac{T}{5000})^{0.46}  \;\textrm{s}^{-1} \nonumber \\
D^{(2)}(5) &=& 1.11 \times 10^{-9} \; n_H (\frac{T}{5000})^{0.44}  \;\textrm{s}^{-1} \nonumber
\end{eqnarray} 

 \subsection{Population and alignment transfer rates: upper levels}
\begin{eqnarray} 
 C_I^{(0)}(2, 1) &=& 8.54 \times 10^{-9} \; n_H (\frac{T}{5000})^{0.38}  \;\textrm{s}^{-1} \nonumber \\
C_I^{(2)}(2, 1) &=& 0.54 \times 10^{-9} \; n_H (\frac{T}{5000})^{0.57}  \;\textrm{s}^{-1} \nonumber \\
C_I^{(0)}(3, 1) &=& 1.51 \times 10^{-9} \; n_H (\frac{T}{5000})^{0.46}  \;\textrm{s}^{-1} \nonumber \\
C_I^{(2)}(3, 1) &=& 1.37 \times 10^{-9} \; n_H (\frac{T}{5000})^{0.48}  \;\textrm{s}^{-1} \nonumber \\
C_I^{(0)}(3, 2) &=& 1.90 \times 10^{-9} \; n_H (\frac{T}{5000})^{0.42}  \;\textrm{s}^{-1} \nonumber \\
C_I^{(2)}(3, 2) &=& 0.65 \times 10^{-9} \; n_H (\frac{T}{5000})^{0.53}  \;\textrm{s}^{-1} \nonumber \\
C_I^{(0)}(4, 2) &=& 1.16 \times 10^{-9} \; n_H (\frac{T}{5000})^{0.47}  \;\textrm{s}^{-1} \nonumber \\
C_I^{(2)}(4, 2) &=& 1.15 \times 10^{-9} \; n_H (\frac{T}{5000})^{0.47}  \;\textrm{s}^{-1} \nonumber \\
C_I^{(0)}(4, 3) &=& 1.72 \times 10^{-9} \; n_H (\frac{T}{5000})^{0.42}  \;\textrm{s}^{-1} \nonumber \\
C_I^{(2)}(4, 3) &=& 0.99 \times 10^{-9} \; n_H (\frac{T}{5000})^{0.47}  \;\textrm{s}^{-1} \nonumber \\
C_I^{(0)}(5, 3) &=& 0.93 \times 10^{-9} \; n_H (\frac{T}{5000})^{0.49}  \;\textrm{s}^{-1} \nonumber \\
C_I^{(2)}(5, 3) &=& 0.94 \times 10^{-9} \; n_H (\frac{T}{5000})^{0.48}  \;\textrm{s}^{-1} \nonumber \\
C_I^{(0)}(5, 4) &=& 1.40 \times 10^{-9} \; n_H (\frac{T}{5000})^{0.43}  \;\textrm{s}^{-1} \nonumber \\
C_I^{(2)}(5, 4) &=& 1.02 \times 10^{-9} \; n_H (\frac{T}{5000})^{0.46}  \;\textrm{s}^{-1} \nonumber 
\end{eqnarray} 

\section{Ionized calcium}  

With obvious notation, these are the collisional rates for 
the Ca {\sc ii} levels of Fig. 5:

 \subsection{Elastic collisional rates}
 \begin{eqnarray} 
 D^{(2)}(P_{3/2}) &=& 5.20 \times 10^{-9} \; n_H (\frac{T}{5000})^{0.41}  \;\textrm{s}^{-1} \nonumber \\  
 D^{(2)}(D_{3/2}) &=& 1.99 \times 10^{-9} \; n_H (\frac{T}{5000})^{0.40}   \;\textrm{s}^{-1}  \\ 
D^{(2)}(D_{5/2}) &=& 2.11 \times 10^{-9} \; n_H (\frac{T}{5000})^{0.38}  \;\textrm{s}^{-1} \nonumber 
\end{eqnarray} 

\subsection{Population and alignment transfer rates}
 
\begin{eqnarray} 
C_I^{(0)}(P_{3/2}, P_{1/2}) &=& 5.70 \times 10^{-9} \; n_H (\frac{T}{5000})^{0.41}  \;\textrm{s}^{-1} \nonumber \\
C_I^{(0)}(D_{5/2}, D_{3/2}) &=& 2.21 \times 10^{-9} \; n_H (\frac{T}{5000})^{0.39}  \;\textrm{s}^{-1} \nonumber \\
C_I^{(2)}(D_{5/2}, D_{3/2}) &=& 1.06 \times 10^{-9} \; n_H (\frac{T}{5000})^{0.49}  \;\textrm{s}^{-1} \nonumber 
\end{eqnarray} 

\end{appendix}


\end{document}